\documentclass[aps,prl,twocolumn,showpacs,superscriptaddress,groupedaddress]{revtex4-1}
\usepackage{amsmath}
\usepackage{graphicx}
\usepackage{bm}        
\usepackage{amssymb}   

\makeatletter
\usepackage{lipsum}
\usepackage{color}
\usepackage{hyperref}
\hypersetup{
     colorlinks   = true,
     citecolor    = blue
}

\makeatother

\begin{document}

\title{\noindent Edge states in a ferromagnetic honeycomb lattice with armchair
boundaries}

\author{\noindent Pierre A. Pantale{\'o}n}
\email{ppantaleon@uabc.edu.mx}
\affiliation{\noindent Theoretical Physics Division, School of Physics and Astronomy,
University of Manchester, Manchester M13 9PL, United Kingdom}

\author{\noindent Y. Xian}
\email{yang.xian@manchester.ac.uk}
\affiliation{\noindent Theoretical Physics Division, School of Physics and Astronomy,
University of Manchester, Manchester M13 9PL, United Kingdom}

\begin{abstract}
\noindent {\normalsize{}We investigate the properties of magnon edge
states in a ferromagnetic honeycomb lattice with armchair boundaries.
In contrast with fermionic graphene, we find novel edge states due to the
missing bonds along the boundary sites. After introducing an external
on-site potential at the outermost sites we find that the energy spectra
of the edge states are tunable. Additionally, when a non-trivial gap
is induced, we find that some of the edge states are topologically
protected and also tunable. Our results may explain the origin of
the novel edge states recently observed in photonic lattices. We also
discuss the behavior of these edge states for further experimental
confirmations.}{\normalsize \par}
\end{abstract}
\maketitle
\textit{Introduction.---} One intriguing aspect of electrons moving
in finite-sized honeycomb lattices is the presence of edge states,
which have strong implications in the electronic properties and play
an essential role in the electronic transport \cite{Fujita1996,Nakada1996,Kohmoto2007}.
It is well known that natural graphene exhibits edge states under
some particular boundaries \cite{Wakabayashi2010b,Delplace2011}.
For example, there are flat edge states connecting the two Dirac points
in a lattice with zig-zag \cite{Fujita1996} or bearded edges \cite{Klein1994}.
On the contrary, there are no edge states in a lattice with armchair
boundary \cite{Zhao2012}, unless a boundary potential is applied
\cite{Chiu2012}.

The edge states have also been studied in magnetic insulators \cite{Fujimoto2009,Onose2010,Cao2015},
where the spin moments are carried by magnons. Recently, it has been
shown that the magnonic equivalence for the Kane-Mele-Haldane model
is a ferromagnetic Heisenberg Hamiltonian with the Dzialozinskii-Moriya
interaction \cite{Owerre2016d,Kim2016a}. Firstly, while the energy
band structure of the magnons of ferromagnets on the honeycomb lattice
closely resembles that of the fermionic graphene \cite{Pantaleon2017,Sakaguchi2016},
it is not clear whether or not they show similar edge states, particularly
in view of the interaction terms in the bosonic models which are usually
ignored in graphene \cite{Lado2015}. Secondly, most recent experiments
in photonic lattices have observed novel edge states in honeycomb
lattices with bearded \cite{Plotnik2014} and armchair \cite{Mili2017}
boundaries, which are not present in fermionic graphene. The main
purpose of this paper is to address these two issues. By considering
a ferromagnetic honeycomb lattice with armchair boundaries, we find
that the bosonic nature of the Hamiltonian reveals novel edge states
which are not present in their fermionic counterpart. After introducing
an external on-site potential at the outermost sites, we find that
the edge states are tunable. Interestingly, we find that the nature
of such edge states is Tamm-like \cite{Tamm}, in contrast with the equivalent
model for the armchair graphene \cite{Chiu2012} but, as mentioned
earlier, in agreement with the experiments in photonic lattices \cite{Plotnik2014,Mili2017}.
Furthermore, after introducing a Dzialozinskii-Moriya interaction
(DMI), we find that the topologically protected edge states are sensitive
to the presence of the Tamm-like states and they also become tunable.

\begin{figure}
\begin{centering}
\includegraphics[scale=0.5]{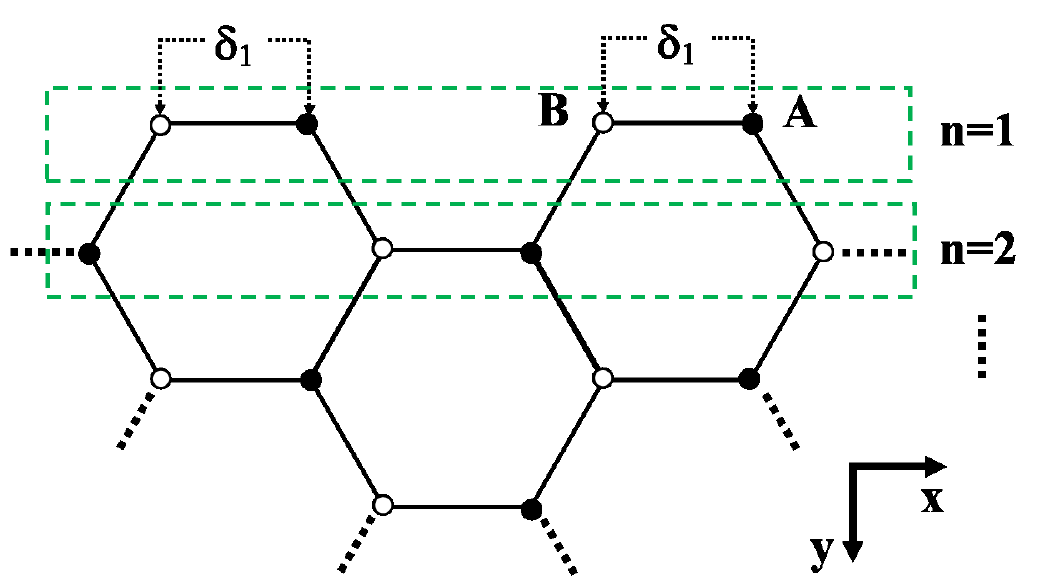}
\par\end{centering}
\caption{{\small{}Squematics of the upper armchair edge of a honeycomb lattice.
The external on-site potential $\delta_{1}$ is applied at the outermost
sites. Here, $n$ is a real-space row index in $y$ direction perpendicular
to the edge. For a large $N$, we consider the opposite edge with
the same structure and with an on-site potential $\delta_{N}$. \label{fig:Figure1}}}
\end{figure}

\textit{Model.---} We consider the following Hamiltonian for a ferromagnetic
honeycomb lattice,

\begin{equation}
H=-J\sum_{\left\langle i,j\right\rangle }\mathbf{S}_{i}\cdot\mathbf{S}_{j}+\sum_{\left\langle \left\langle i,j\right\rangle \right\rangle }\boldsymbol{D}_{ij}\cdot\left(\mathbf{S}_{i}\times\mathbf{S}_{j}\right),\label{eq:HFull}
\end{equation}
where the first summation runs over the nearest-neighbors (NN) and
the second over the next-nearest-neighbors (NNN), $J\left(>0\right)$
is the isotropic ferromagnetic coupling, $\boldsymbol{S}_{i}$ is
the spin moment at site $i$ and $\boldsymbol{D}_{ij}$ is the DMI
vector between NNN sites \cite{Moriya1960}. If we assume that the lattice
is at the $x$-$y$ plane, according to Moriya's rules \cite{Moriya1960},
the DMI vector vanishes for the NN but has non-zero component along
the $z$ direction for the NNN. Hence, we can assume $\boldsymbol{D}_{ij}=D\nu_{ij}\hat{z}$,
where $\nu_{ij}=\pm1$ is an orientation dependent coefficient in
analogy with the Kane-Mele model \cite{Kane2005}. For the infinite
system in the linear spin-wave approximation (LSWA), the Hamiltonian
in Eq. (\ref{eq:HFull}) can be reduced to a bosonic equivalent of
the Kane-Mele-Haldane model \cite{Owerre2016d,Kim2016a,Pantaleon2017}.
To investigate the edge states we consider an armchair boundary along
the $x$ direction, with a large $N$ sites in the $y$ direction,
as shown in Fig. (\ref{fig:Figure1}). A partial Fourier transform
is made and the Hamiltonian given in Eq. (\ref{eq:HFull}) in LSWA
can be written in the form,
\begin{equation}
H=-t\sum_{k}\Psi_{k}^{\dagger}M\Psi_{k},\label{eq: HolePartHamil}
\end{equation}
where $\Psi_{k}^{\dagger}=\left[\Psi_{k,A}^{\dagger},\,\Psi_{k,B}^{\dagger}\right]$
is a $2N\times2N$, $2$-component spinor, $k$ is the Bloch wave
number in the $x$ direction and $t=JS$. The matrix elements of $M$
are $N\times N$ matrices given by: $M_{11}=\left(1-\delta_{1}\right)T^{\dagger}T+\left(1-\delta_{N}\right)TT^{\dagger}+\left(1+\delta_{1}+\delta_{N}\right)I+M_{D}$,
$M_{12}=-J_{1}I-J_{2}\left(T+T^{\dagger}\right)$, $M_{21}=M_{12}^{\dagger}$,
$M_{22}=M_{11}-2M_{D}$ and $M_{D}=J_{3}\left(TT-T^{\dagger}T^{\dagger}\right)+J_{4}\left(T+T^{\dagger}\right)$
the DMI contribution. Here, $T$ is a displacement matrix as defined
in Ref. \cite{You2008a} and $I$ a $N\times N$ identity matrix.
We have also introduced two on-site energies $\delta_{1}$ and $\delta_{N}$
at the outermost sites of each boundary, respectively. The coupling
terms are: $J_{1}=e^{-ik}$, $J_{2}=e^{\frac{1}{2}ik}$, $J_{3}=i\,D^{\prime}$,
$J_{4}=2i\,D^{\prime}\,\cos\left(3k/2\right)$ and $D^{\prime}=D/J$.
The numerical diagonalization of the matrix given by Eq. (\ref{eq: HolePartHamil})
reveals that the bulk spectra is gapless only if $N=3m+1$, with $m$
a positive integer \cite{Wang2016d}. However, to avoid size-dependent
bulk gaps or hybridization between edge states of opposite edges \cite{Chiu2012},
we consider a large $N$ where the edge states are independent of
the size \cite{Hatsugai1993,Hatsugai1993a}.

\textit{Boundary conditions.---} From the explicit form of the matrix
elements in Eq. (\ref{eq: HolePartHamil}), the coupled Harper equations
can be obtained \cite{Harper1955}. If we assume that the edge states
are exponentially decaying from the armchair boundary, we can consider
the following anzats \cite{Konig2008,Wang2009} for the eigenstates
of $M$ in Eq. (\ref{eq: HolePartHamil}),
\begin{equation}
\Psi_{k}\left(n\right)=\left[\begin{array}{c}
\psi_{k,A}\left(n\right)\\
\psi_{k,B}\left(n\right)
\end{array}\right]=z^{n}\left[\begin{array}{c}
\phi_{k,A}\\
\phi_{k,B}
\end{array}\right],\label{eq: Main Anzats}
\end{equation}
where $\left[\phi_{k,A},\,\phi_{k,B}\right]^{t}$ is an eigenvector
of $M$, $z$ is a complex number and $n\,\left\{ =1,\,2,\,3\,...\right\} $
is a real space lattice index in the $y$ direction, as shown in Fig.
(\ref{fig:Figure1}). Upon substitution of the anzats in the coupled
Harper equations, the complex number $z$ obey the following polynomial
equation,
\begin{equation}
\sum_{\mu=0}^{4}a_{\mu}(z+z^{-1})^{\mu}=0,\label{eq: ArmchairCharacteristic}
\end{equation}
with coefficients: $a_{0}=1-\left(3-\varepsilon\right)^{2}-4J_{4}^{2}$,
$a_{1}=8J_{3}J_{4}+J_{1}^{\ast}J_{2}+J_{2}^{\ast}J_{1}$, $a_{2}=-4J_{3}^{2}+J_{4}^{2}+1$,
$a_{3}=-2J_{3}J_{4}$ and $a_{4}=J_{3}^{2}$. For a given $k$ and
energy $\varepsilon$, such a polynomial always yields four solutions
for $\left(z+z^{-1}\right)$. Since we require a decaying wave from
the boundary, only the solutions with $\left|z\right|<1$ are relevant
for the description of the edge states at the upper edge and $\left|z\right|>1$
for the lower (opposite) edge. The eigenfunction of Eq. (\ref{eq: HolePartHamil})
satisfying $\underset{n\rightarrow\infty}{lim}\,\Psi_{k}\left(n\right)=0$
may now in general be written as,
\begin{equation}
\psi_{k,l}\left(n\right)=\sum_{\upsilon=1}^{4}c_{v}z_{v}^{n}\phi_{l,v},\label{eq: ArmchaLinComb}
\end{equation}
where the coefficients $c_{\upsilon}$ are determined by the boundary conditions and
$\phi_{l,v}$ is the two-component eigenvector $\left(l=A,\,B\right)$
of $M$. From the Harper equations provided by the Eq. (\ref{eq: HolePartHamil})
and Eq. (\ref{eq: Main Anzats}), the boundary conditions are satisfied
by,
\begin{align}
\left(1-\delta_{1}\right)\psi_{k,A}\left(1\right)-J_{2}\psi_{k,B}\left(0\right) & =0,\label{eq:ArmBound1}\\
\left(1-\delta_{1}\right)\psi_{k,B}\left(1\right)-J_{2}^{\ast}\psi_{k,A}\left(0\right) & =0,\label{eq: ArmBound2}\\
J_{4}\psi_{k,A}\left(0\right)-J_{3}\psi_{k,A}\left(-1\right) & =0,\label{eq: ArmBound3}\\
J_{4}\psi_{k,B}\left(0\right)-J_{3}\psi_{k,B}\left(-1\right) & =0.\label{eq: ArmBound4}
\end{align}
By Eq. (\ref{eq: ArmchaLinComb}), the above relations can be written
as a set of equations for the unknown coefficients $c_{v}$. The non-trivial
solution and the polynomial given by Eq. (\ref{eq: ArmchairCharacteristic}),
provide us a complete set of equations for the edge state energy dispersion
and they can be solved numerically. The same procedure can be followed
to obtain the solutions for the opposite edge.
\begin{figure}
\begin{centering}
\includegraphics[scale=0.44]{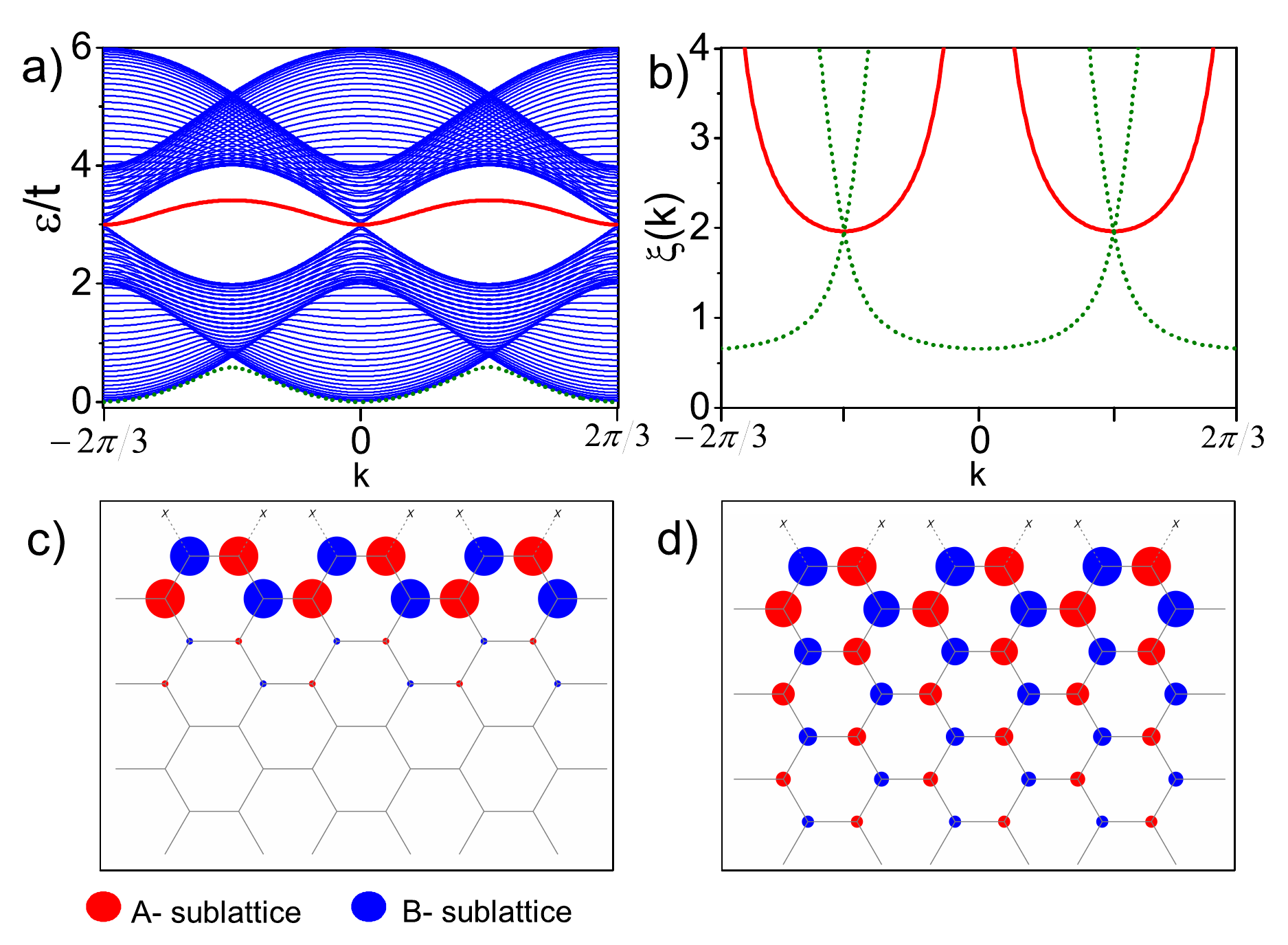}
\par\end{centering}
\caption{{\small{}(Color online) a) Edge state energy dispersion for $D=0$
and $\delta_{1}=\delta_{N}=0$. The blue regions are the bulk energy
spectra. The green (dotted) and red (continuous) lines are the edge
state energy bands. In b) their corresponding penetration lengths
are shown. The magnon density profile for the edge magnon is shown
in c) for $\varepsilon=\left(2\pm\sqrt{2}\right)t$ at $k=\pm\pi/3$,
and in d) for $\varepsilon=0.298\,t$ at $k=\pm0.65$. The radius
of each circle is proportional to the magnon density.\label{fig: Figure2}}}
\end{figure}

\textit{Zero DMI.---} For the system without DMI, the coupling terms
invoving $J_{3}$ and $J_{4}$ vanish, and the boundary conditions
are reduced to the Eqs. (\ref{eq:ArmBound1}) and (\ref{eq: ArmBound2})
with a quadratic polynomial in $(z+z^{-1})$ of Eq. (\ref{eq: ArmchairCharacteristic}).
In particular, for the (uniform) case with $\delta_{1}=\delta_{N}=1$,
the edge and the bulk sites have the same on-site potential and the
boundary conditions provide us with two bulk solutions with $z^{2}=1$.
Therefore, in analogy with graphene with armchair edges, there are
not edge states \cite{Zhao2012}. However, as shown in Fig. (\ref{fig: Figure2}a),
in the absence of external on-site potential $\left(\delta_{1}=\delta_{N}=0\right)$,
two new dispersive localized modes are obtained. Located between (red,
continuous line) and below (green, dotted) the bulk bands, such edge
states are well defined along the Brillouin zone and their energy
bands are doubly degenerated due to the fact that there are two edges
in the ribbon.

These edge states have not been previously predicted or observed in
magnetic insulators. However, we believe that they are analogous to
the novel edge states recently observed in a photonic honeycomb lattice
with armchair edges \cite{Mili2017}. Although in Ref. \cite{Mili2017}
these edge states may be attributed to the dangling bonds along the
boundary sites (the details have been given for zig-zag and bearded
but not for armchair edges), and since these dangling bonds can be
viewed as effective defects along the edges, similar physics is contained
in our model where the effective defects are described by the different
on-site potential at the boundaries. We believe that our approach
has the advantage of simple implementation for various boundary conditions.
In particular, we have obtained analytical expressions for the wavefunctions
and their confinement along the boundary. The latter is given by the
penetration length (or width) of the edge state \cite{Doh2013} defined
as,
\begin{equation}
\xi_{i}\left(k\right)\equiv\frac{\sqrt{3}}{2}\left[\ln\left|\frac{1}{z_{i}\left(k\right)}\right|\right]^{-1},\label{eq: PenetrationLength}
\end{equation}
indicating a decay of the form $\sim e^{-y/\xi_{i}\left(k\right)}$.
In the above equation, $z_{i}$ is the $i$-th decaying factor in
the linear combination, Eq. (\ref{eq: ArmchaLinComb}). Since we require
two decaying factors to construct the edge state, we have two penetration
lengths as mentioned in Ref. \cite{Mili2017}. The penetration lengths
for the edge states with $\delta_{1}=\delta_{N}=0$ are shown in the
Fig. (\ref{fig: Figure2}b). The dotted (green) and continuous (red)
lines are the corresponding penetration lengths for the edge states
in the Fig. (\ref{fig: Figure2}a). The edge state between the bulk
bands (red, continuous) is composed by two penetration lengths but
they are indistinguishable to each other. This indicates that the
decaying factors are complex conjugates to each other, $z_{1}=z_{2}^{\ast}$
\cite{Doh2013,Pantaleon2017}. The edge state below the lower bulk
band (green, dotted) depends on two penetration lengths in the region
around $k=\pm\pi/3$. However, outside such region, one penetration
length diverges, $\left|z\right|\rightarrow1$, while the other one
decreases to a minimum value. This means that the edge state tends
to merge with the bulk and is almost indistinguishable at $k=0,\,\pm2\pi/3$.
Furthermore, as we can see in the Fig. (\ref{fig: Figure2}b), at
$k=\pm\pi/3$, the penetration length of both edge states is identical
hence they have their maximum confinement along the boundary at the
same Bloch wave-vector. This is shown in the Fig. (\ref{fig: Figure2}c)
where we plot the magnon density, $\left|\psi_{k,l}\left(n\right)/\psi_{k,l}\left(1\right)\right|^{2}$,
for both edge states at $k=\pm\pi/3$ with their corresponding energies,
$\varepsilon=\left(2\pm\sqrt{2}\right)t$. In addition, in the Fig.
(\ref{fig: Figure2}d) the magnon density for the edge state below the
lower bulk band with energy $\varepsilon=0.298\,t$ at $k=\pm0.65$ is shown, where as we mentioned before,
the edge state tends to spread to the inner sites.

\begin{figure}
\begin{centering}
\includegraphics[scale=0.17]{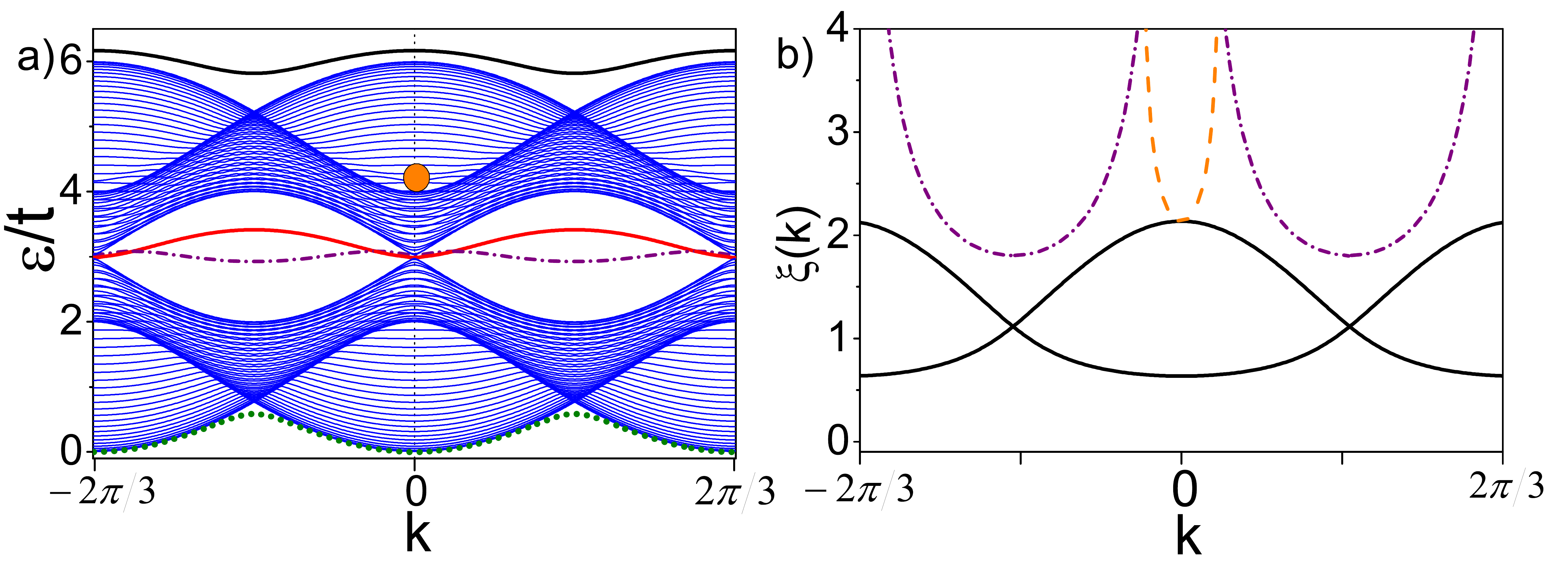}
\par\end{centering}
\caption{{\small{}(Color online) (Color online) a) Edge state energy dispersion
for $D=0$ with $\delta_{1}=2.5$ and $\delta_{N}=0$. The blue regions
are the bulk energy spectra. The dotted (green) and continuous (red)
lines are the edge state energy bands at the lower edge. The dot-dashed
(purple) line, continuous (black) line and the circle (orange) are
the edge states at the upper edge. In b) the penetration lengths of
the corresponding edge states at the upper edge is shown. \label{fig: Figure3}}}
\end{figure}

The edge states discussed above have been obtained with no gap in
the bulk and their dispersion relations are between the Dirac points.
Their existence without external on-site potentials indicates that
the edge states are ``Tamm-like'' \cite{Tamm,Plotnik2014}. Such
type of states are usually associated with surface perturbations or
defects. However, in our system no defects are present. The nature
of such edge states can be explained in terms of the intrinsic on-site
potential, where each site along the armchair boundary has two nearest-neighbors
and the intrinsic on-site potential is lower than in the bulk. Such
a difference plays the same role as an effective defect which allows
the existence of a Tamm-like state.
\begin{figure*}[t]
\begin{centering}
\includegraphics[scale=0.15]{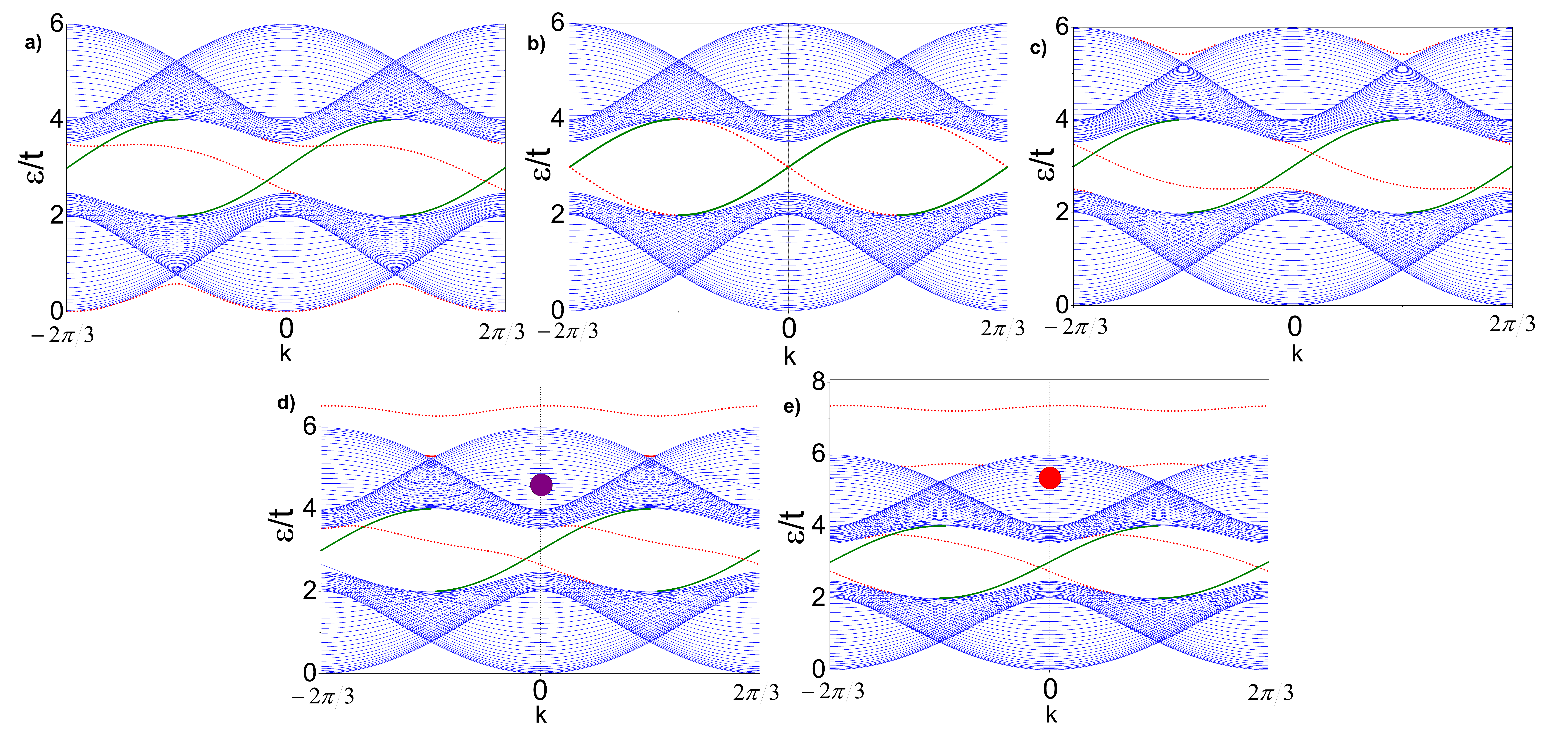}
\par\end{centering}
\caption{{\small{}(Color online) Edge state energy dispersion for $D=0.1\,J$.
The blue regions are the bulk energy spectra. For $\delta_{N}=1$
(lower edge) there is an edge state crossing the gap (green, continuous).
The red (dotted) lines are the edge states for a) $\delta_{1}=0$,
b) $\delta_{1}=1$, c) $\delta_{1}=2$, d) $\delta_{1}=3$ and e)
$\delta_{1}=4$. The circles in d) and e) are edge states within the
bulk energy bands. \label{fig:Figure4}}}
\end{figure*}

We next discuss the effects of edge potentials. It has been shown
that edge states can be induced by edge potentials in the armchair
graphene \cite{Chiu2012}, however, the edge states that we found
are a consequence of the bosonic nature of the lattice as discussed
above. Since they exist without opening a gap and they are dispersive,
it is not clear if they can be predicted by a topological approach.
Our approach reveals that the edge states in a bosonic lattice are
strongly dependent on the on-site potential along the boundary. For
example, if $\delta_{1}=2.5$ (with $\delta_{N}=0$) some new features
are obtained. As shown in Fig. (\ref{fig: Figure3}a), the presence
of the strong external on-site potential reveals three edge states at the upper boundary:
a high energy edge state over the bulk bands (black, continuous line),
an edge state between the bulk bands (purple, dot-dashed line), and
interestingly, an edge state within the bulk bands (orange circle).
Such edge state is strongly localized and is highly dispersive. It
merges into the bulk with an small change in their Bloch wave-number
and may be difficult to detect in a magnetic insulator. It is therefore
very encouraging that similar edge state was observed in a photonic
lattice \cite{Mili2017}. The penetration lengths of the corresponding
edge states are shown in the Fig. (\ref{fig: Figure3}b), firstly,
the edge state over the bulk bands (black, continuous line) depends on two  decaying factors and the oscillating behavior of their corresponding penetration lengths reveals that is strongly localized along the boundary sites and it never merges into the bulk. Secondly, the edge state within the bulk bands (orange,
dashed line) depends on a single decaying factor and is highly dispersive.
Thirdly, the edge state between the bulk bands (purple, dot-dashed)
depends on a single penetration length since the two decaying factors
are conjugate to each other \cite{Doh2013}. 

\textit{Non-zero DMI.---} It is well known that a non-zero DMI in a
bosonic honeycomb lattice makes the band structure topologically non-trivial
and reveals metallic edge states which transverse the gap \cite{Owerre2016d}.
However, the edge states that appear, under, within and over the bulk
bands in Fig. (\ref{fig: Figure2}a) and Fig. (\ref{fig: Figure3}a)
are distinct to the edge states predicted by topological arguments.

In the Fig. (\ref{fig:Figure4}) we show the energy bands for a DMI
strength of $D=0.1\,J$, where we keep a fixed $\delta_{N}=1$ and
we modified $\delta_{1}$. The continuous (green) line that cross
the gap from the lower to the upper bulk bands is the edge state at
the lower edge. The dotted (red) lines correspond to the edge states
at the upper edge. If we follow the edge state energy
spectra at the upper boundary from the Fig. (\ref{fig:Figure4}a)
to the Fig. (\ref{fig:Figure4}c), we observe that the edge state
within the bulk gap change its concavity. The Tamm-like state below
the bulk bands, Fig. (\ref{fig:Figure4}a), merge with the bulk and
a new Tamm-like state appears at the top of the upper bulk band, as
shown in Fig. (\ref{fig:Figure4}c). If we keep increasing the value
of the external on-site potential the Tamm-like state over the bulk
band in Fig. (\ref{fig:Figure4}c) moves away from the upper bulk
band, Fig. (\ref{fig:Figure4}d). Furthermore, a second Tamm-like
state appears with components within the bulk, as shown in Fig. (\ref{fig:Figure4}d)
and Fig. (\ref{fig:Figure4}e). The boundary conditions suggest that
the existence of these two tunable edge states is due to the two sites
in the unit cell of the armchair boundary and, by symmetry, the same
behavior is expected at the opposite edge. These edge states can be
made to locate below, within and over the bulk bands. If a non-trivial
gap is induced the topologically protected edge states are also tunable.

Finally, a similar phenomena is expected for a lattice with zig-zag
or bearded boundaries. In both cases there is a single outermost site
and Tamm-like edge states may appear due to the missing bond and/or
by the external on-site potential. Since the outermost site at the
lattice with a bearded boundary has three missing bonds, the effective
defect should be stronger than the corresponding to a zig-zag boundary.
This may be related with the existence of the ``unconventional''
edge states found in optical lattices \cite{Plotnik2014}. A more
extensive investigation of Tamm-like edge states along different boundaries will be reported elsewhere.

\textit{Conclusions.---} We have analyzed the edge states in a ferromagnetic
honeycomb lattice with armchair edges and an external on-site potential
at the outermost sites. In contrast with graphene, our system without
external on-site potential reveals two edge states. It is clear that
the open boundary in a bosonic lattice creates an effective defect
by a difference in the on-site potential between the bulk and boundary
sites. This effective defect is responsible for the existence of the
novel edge states. By introducing an external on-site potential at
the outermost sites we found that the nature of this edge states
is Tamm-like. We also found that these edge states are tunable in
their shapes and positions depending on the external on-site potential
strength. Such tunability can be used to modify the topologically
protected edge states when a non-trivial gap is induced. Finally,
we found that the number of these tunable edge states is related
to the number of sites in a unit cell along the boundary. We believe
that our results may explain the edge states recently found
in optical lattices \cite{Plotnik2014,Mili2017} and motivate new
experiments in both magnonic and photonic lattices.

Pierre. A. Pantale{\'o}n is sponsored by Mexico's National
Council of Science and Technology (CONACYT) under the scholarship
381939.

\bibliographystyle{apsrev4-1}

\end{document}